\documentclass[a4paper,11pt]{article}
\pdfoutput=1 

\usepackage{jheppub} 

\usepackage[T1]{fontenc} 

\usepackage{graphicx}
\usepackage{dcolumn}
\usepackage{color}
\usepackage[T1]{fontenc}
\usepackage{amssymb,amsmath}
\usepackage{ulem}
\usepackage{mathrsfs}

\newcommand{\bfx}{\mbox{\boldmath $x$}}

\newcommand{\bfp}{\mbox{\boldmath $p$}}

\newcommand{\bfk}{\mbox{\boldmath $k$}}
\def\be{\begin{equation}}
\def\ee{\end{equation}}
\def\ba{\begin{eqnarray}}
\def\ea{\end{eqnarray}}

\title{\boldmath Conversions of propagation eigenstates of supernova neutrinos by atomic electrons}

\author[a,1]{Motohiko Kusakabe\note{Corresponding author.}}

\affiliation[a]{School of Physics, and
  International Research Center for Big-Bang Cosmology and Element Genesis,
  Beihang University \\
  37, Xueyuan Rd., Haidian-qu, Beijing 100083 China}

\emailAdd{kusakabe@buaa.edu}

\abstract{
  Electron number densities in stars and the Earth are inhomogeneous because of atomic electrons. The large inhomogeneities on atomic-scale tend to form at tops of respective layers of stars, and 1s electrons of O locally produce weak potentials higher than that of the high MSW resonance. Then, supernova neutrinos experience vast numbers of non-adiabatic transitions. This inhomogeneous electron potential generates finite amplitudes of all three propagation eigenstates, and wave packets effectively separate. Then, spectral differences between three flavors significantly diminishes after propagation.
}

\begin{document} 
\maketitle
\flushbottom

\section{Introduction}
\label{sec1}

Neutrino flavor eigenstates are described as superpositions of mass eigenstates \cite{1962PThPh..28..675K,Maki:1962mu}. As neutrinos propagate, their flavors change between three eigenstates \cite{Pontecorvo:1957cp,Pontecorvo:1967fh} due to different eigenvalues. The flavor eigenstates $|\nu_\alpha\rangle$ for $\alpha =e$, $\mu$, $\tau$ are related to the mass eigenstates $| n_i \rangle$ for $i=1$--3, via
$| \nu_\alpha \rangle = U_{\alpha i}^\ast | n_i \rangle$, where the unitary matrix $U_{\alpha i}$ is called Pontecorvo-Maki-Nakagawa-Sakata matrix.

When neutrinos propagate in matter, flavor evolution differs from the vacuum case \cite{Wolfenstein:1977ue,Mikheev:1986gs,Mikheev:1987qk}.
The Mikheyev-Smirnov-Wolfenstein (MSW) resonances occur at electron number densities of
\ba
n_{e,{\rm res}} =\frac{1}{2 \sqrt{2} G_{\rm F}} \frac{\Delta m^2}{E} \cos(2 \theta),
\ea
where $G_{\rm F}$ is the Fermi constant,
$\Delta m^2$ is a mass squared difference,
$\theta$ is a neutrino mixing angle, and
$E$ is the neutrino energy.
A resonance for mass eigenstates 1 and 2 exists at a low density, i.e., L-resonance:
$n_{e,{\rm L}} = 1.15 \times 10^{25}~[E/(10~{\rm MeV})]^{-1}$ cm$^{-3}$.
A resonance for states 1 and 3 is at a high density, i.e., H-resonance:
$n_{e,{\rm H}} = 9.54 \times 10^{26}~[E/(10~{\rm MeV})]^{-1}$ cm$^{-3}$.
The states 2 and 3 mix almost completely, and there is no resonance.

Large-scale random perturbations of matter density in the MSW region of the Sun affect neutrino oscillations if the perturbation scale is close to the oscillation length in matter \cite{Nunokawa:1996qu}. However, helioseismic waves are much smaller than the density perturbation required for observable effects on the neutrino spectrum \cite{Bamert:1997jj}.

In this paper, we show the existence of inhomogeneity in the electron number density $n_e$ in stars caused by atomic electrons. Supernova (SN) neutrinos experience many non-adiabatic transitions of propagation eigenstates, and wave packet (WP) separations result in effective flavor changes.
We use the natural units, i.e., $\hbar=c=k_{\rm B}=1$.
Neutrino parameters are adopted from Ref. \cite{Zyla:2020zbs}:
$\sin^2 \theta_{12} =0.307$, 
$\Delta m^2_{21} =7.53 \times 10^{-5}$ eV$^2$,
$\sin^2 \theta_{23} =0.545$, 
$\Delta m^2_{32} =2.453 \times 10^{-3}$ eV$^2$ (if normal ordering [NO]) or $-2.546 \times 10^{-3}$ eV$^2$ (if inverted ordering),
$\sin^2 \theta_{13} =0.0218$, and 
CP violating phase $\delta /\pi =1.36$.
The sign of $\Delta m^2_{32} =m^2_{3} -m^2_{2}$ is still undetermined.
In this paper, the NO is assumed and only neutrinos are treated.

\section{Stable existence of 1s electrons}\label{sec2}

The adiabaticity parameter \cite{1932PhyZS...2...46L,Zener:1932ws,Fukugita:2003en} is given by
\ba
\gamma_{ij} &=& \frac{\Delta m^2_{ji} \sin^2(2 \theta_{ij})}{2 E \cos(2 \theta_{ij}) d \ln n_e /dr}.
\ea
We adopt the published pre-SN data \cite{Kusakabe:2019znq} for the progenitor of SN 1987 A based upon a Kyushu method \cite{Kikuchi:2015ena}.
For the density $\rho =10^3$ g cm$^{-3}$, the corresponding radius $r \sim 6 \times 10^9$ cm, and $d \ln n_e /dr \sim 1/(10^4~{\rm km})$ for the $n_e$ averaged over large volume, i.e., $\langle n_e \rangle$ \cite{Kusakabe:2019znq}.
Then, $\gamma_{23} =-6.85 \times 10^4 ({E}/{10~{\rm MeV}})^{-1}$.
Thus, the adiabaticity is very high for the $\langle n_e \rangle$ profile.

The size of electronic 1s orbital of element with atomic number $Z$ roughly scales as
$a_B/Z \approx 6.61 \times 10^{-10} (Z/8)^{-1}$ cm, where $a_B$ is the Bohr radius. The atomic electron causes a density gradient 
$|d \ln n_e /dr| \sim 1.51 \times 10^{9}~{\rm cm}^{-1} (Z/8)$, and when a neutrino passes through an atom the adiabaticity is $\gamma_{23} =-4.53 \times 10^{-14} (Z/8)^{-1}$.
This is extremely small.
The difference between the homogeneous (simplified) and inhomogeneous (realistic) $n_e$ profiles of SN are then adiabatic versus non-adiabatic transitions.


For pure composition of an element $i$ with $Z$, the Saha equation tells that the ionization degree $\chi_i$ of $i^{(Z-2)+}$ satisfies
\be
\frac{1-\chi_i}{\chi_i} =\rho N_A Y_e f_e^{\rm ion}
\left( \frac{m_e T}{2 \pi} \right)^{-3/2} \exp \left( \frac{B_i}{T} \right),
\ee
where $N_A$ is the Avogadro number, $Y_i$ is the mole fraction of species $i$,
$f_e^{\rm ion} =[(Z-2) +\chi_i]/Z$ is the free electron fraction, 
$m_e$ is the electron mass, $T$ is the temperature,
and $B_i$ is the electron binding energy, which is much less than $T$ deep inside stars.
Inside stars, recombinations are relatively strong, and the ionization degree is tiny, i.e., $\chi_i \ll 1$.
In the He and O-Ne-Mg layers in pre-SN stars, almost all electrons are in bound states as far as those states can exist stably (i.e., the cutoff in the partition function \cite{1979rpa..book.....R}).
Supposing that compositions of O-Ne-Mg layer are dominated by O and only 1s electrons form stable bound states (see below),
approximately $f_e^{\rm ion} \approx (6 +\chi_{\rm O})/8$.
Then, it follows that
\be
\frac{1-\chi_{\rm O}}{\chi_{\rm O}} \approx 1.7\times 10^{21} \left( \frac{\rho}{10^3~{\rm g~cm}^{-3}} \right) \left( \frac{Y_e f_e^{\rm ion}}{3/8} \right)
\left( \frac{T}{10^9~{\rm K}} \right)^{-3/2}.
\ee
The ionization degree is then $\chi_{\rm O} \ll 1$. Similarly, the second 1s electron is completely bound to O in stars.

Table \ref{tab1} lists the critical density satisfying
$a_{\rm B}/Z = [ 3/(4 \pi \rho_{\rm cr} N_A Y_e)]^{1/3}$.
The typical distance to the nearest electron is shorter than the orbital radius if $\rho > \rho_{\rm cr}$, where the 1s state cannot exist stably.
The third column shows the maximum 1s electron number densities in respective atoms.
$_6$C, $_8$O, and heavier elements have central $n_e$ values above the H resonance density.

\begin{table}[t]
  \caption{\label{tab1}%
    The critical density for stability of 1s states and the maximum electron number density at atomic center.}
\begin{tabular}{lD{.}{.}{5}D{.}{.}{5}}
  \hline
  element & \multicolumn{1}{c}{$\rho_{\rm cr}$ (10$^3$ g cm$^{-3}$)} & \multicolumn{1}{c}{$\Delta n_{e}$ ($10^{26}$ cm$^{-3}$)} \\
  \hline
  H  & 0.00268 Y_e^{-1}       & 0.04296 \\
  He & 0.0428 (Y_e /0.5)^{-1} & 0.2064 \\
  C  & 1.16 (Y_e /0.5)^{-1}   & 7.904 \\
  O  & 2.74 (Y_e /0.5)^{-1}   & 19.52 \\
  \hline
\end{tabular}
\end{table}

%

\section{Conversions of propagation eigenstates}\label{sec3}
\begin{figure}[tbp]
\begin{center}
\includegraphics[width=0.8\linewidth]{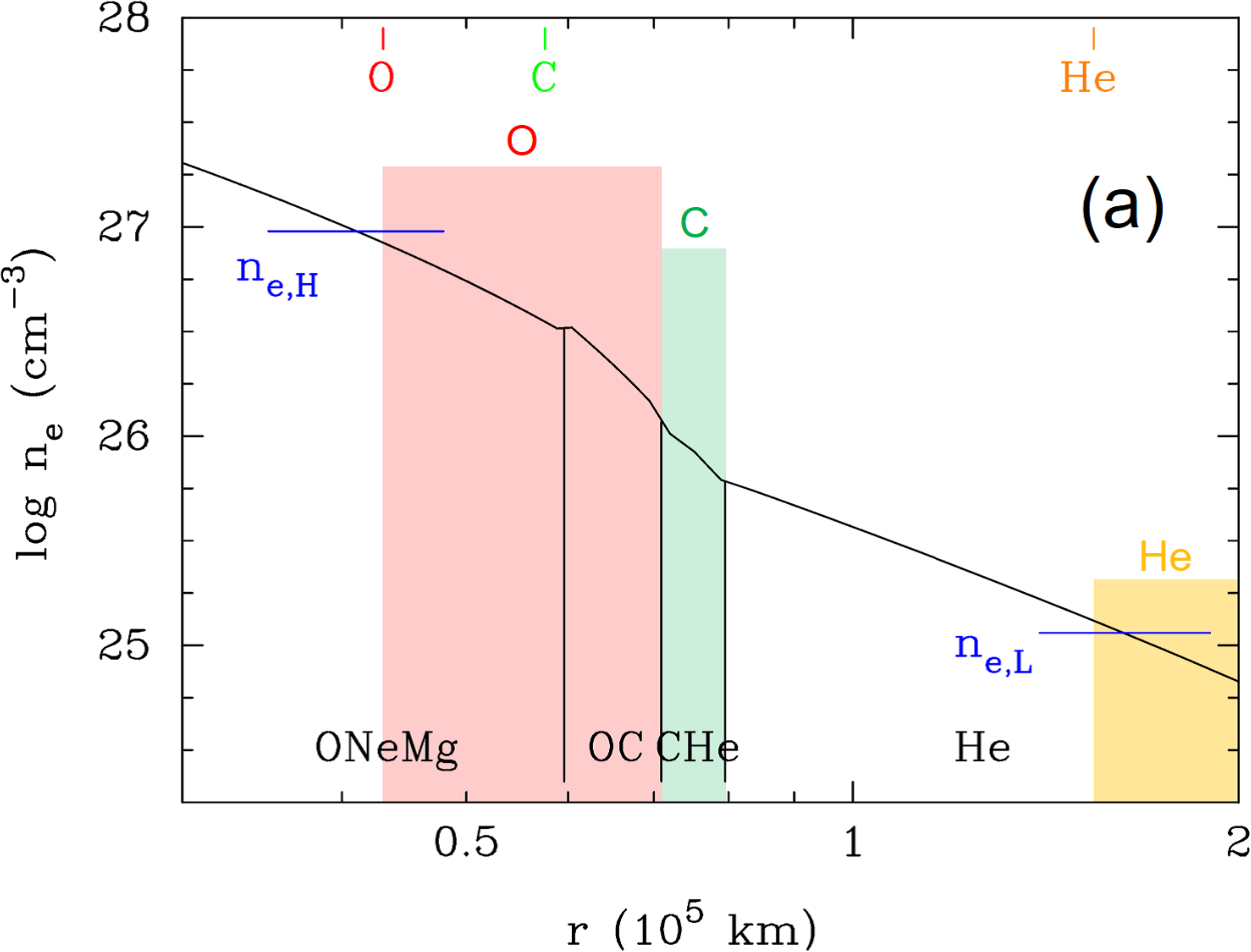}
\includegraphics[width=0.8\linewidth]{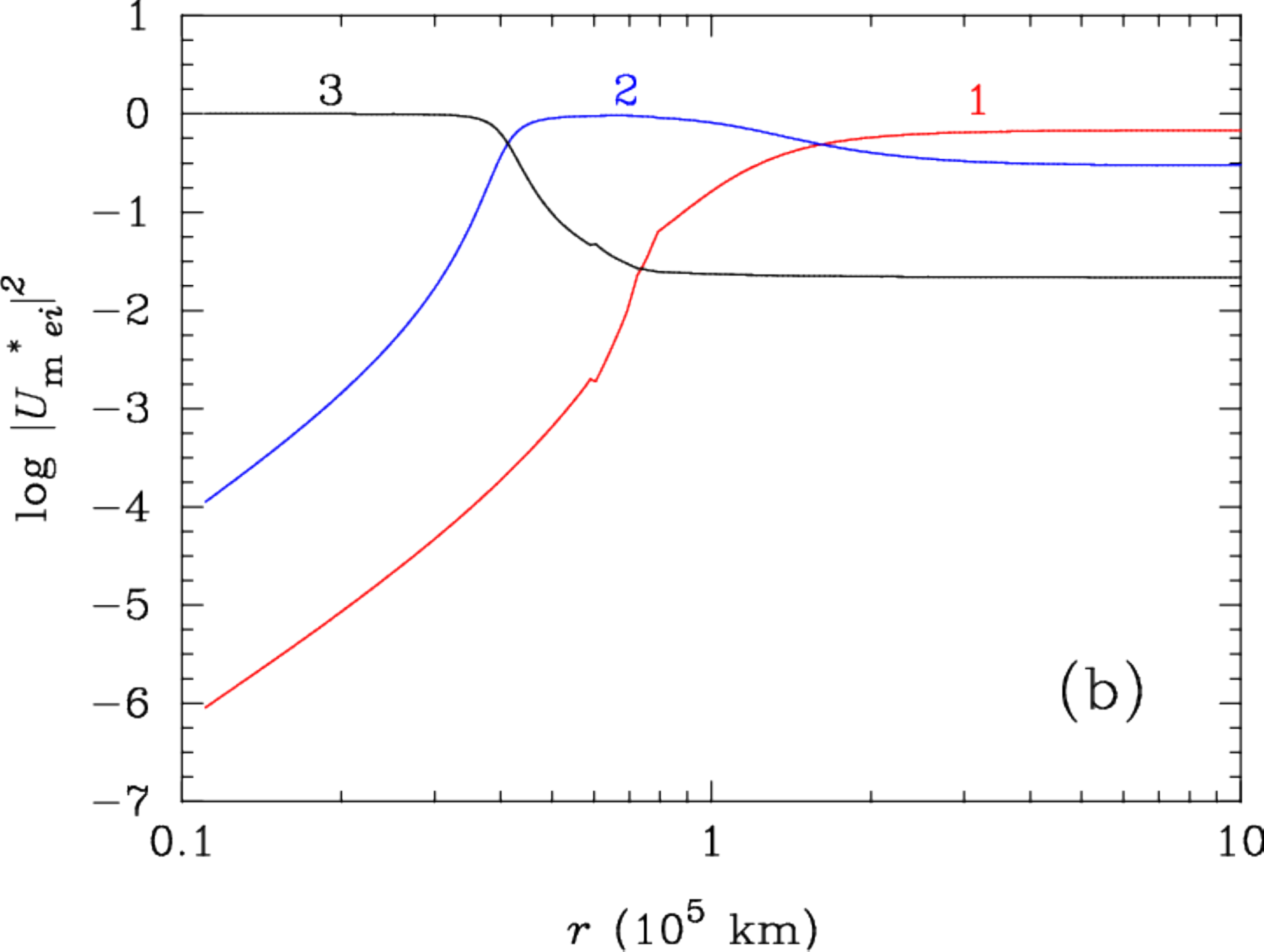} 
\caption{(a) The profile of average electron number density $n_e$ as a function of radius from the center (solid curve) in the pre-SN model for SN 1987A \cite{Kusakabe:2019znq}. The vertical lines divide layers with different compositions, i.e., O-Ne-Mg, O-C, C-He, and He layers. The three vertical marks on the top identify the locations outside which the 1s electrons of O, C, and He are stabilized. In colored rectangles, the 1s electrons of the most dominant elements cause significant inhomogeneities in $n_e$, and conversions of propagation eigenstates operate. The heights of the colored region identify maximum values of $n_e$ by those atomic electrons. The MSW H- and L-resonances occur for $E=10$ MeV at the horizontal lines.
  (b) The squared amplitudes of effective mixing adjoint elements in homogeneous matter for $\nu_e$, i.e., ${U_{\rm m}^\ast}_{ei}$, for propagation eigenstates $i=1$, 2, and 3 in the NO case.
  \label{fig1_environment}}
\end{center}
\end{figure}

Figure \ref{fig1_environment}(a) shows the average value $\langle n_e \rangle$ versus radius from the center in the pre-SN \cite{Kusakabe:2019znq}.
Outside vertical bars on the top, the 1s electron orbitals of O, C, and He are stabilized, and the most dominant elements make inhomogeneities in $n_e$ via bound electrons in the colored rectangles.
%
Most of electrons distribute homogeneously in the inner region with $\rho >\rho_{\rm cr}$, while they are localized in the orbitals in the outer region.
%
%
The H-resonance density
 is located in the O-rich layer.
In the O-Ne-Mg layer outside the H-resonance region ($\rho \lesssim 10^3$ g cm$^{-3}$), the density is smaller than the critical value for O.
There, electrons are localized in the O 1s state.

Figure \ref{fig1_environment}(b) shows the squared amplitudes for effective mixing adjoint in matter, ${U_{\rm m}^\ast}_{ei}$, versus $r$ for the adopted $\langle n_e \rangle$ value. As a typical SN neutrino energy, $E =10$ MeV is taken.
Matrix operations are performed with Lapack \cite{lapack99}.
Transitions of propagation eigenstates occur adiabatically, and the eigenstates never change under this mean field assumption. Then, always $| \nu_\alpha (t)\rangle = U_{{\rm m}\alpha i}^\ast(t) | n_i (0)\rangle$
with $t=r$ and initial values $| n_i (0) \rangle$.
At the neutrinosphere, a $\nu_e$ is produced as almost pure $n_3$ state. After the propagation, it leaves the star as $\nu_x (\equiv$ a half $\nu_\mu +$ a half $\nu_\tau)$.
Both $\nu_\mu$ and $\nu_\tau$ are produced as a half $n_1$ and a half $n_2$, and evolve to a half $\nu_e$ and a half $\nu_x$.
Thus, flavors change as $e \leftrightarrow x$ by propagation.


%
%
%
%


The transfer equation for propagation eigenstates is \cite{Fantini:2018itu}
\ba
\dot{n}_i(x,t) &=& \left[ -i \Lambda_{\rm m}(x) - U_{\rm m}^T(x) \dot{U}_{\rm m}^\ast(x) \right]_{ij} n_j(x,t)
\label{eq_evo1} \\
d x_i /dt &=&v_{{\rm m},i} =dE_i/dk =1 +d \Lambda_{{\rm m},i}/dk \\
\Lambda_{{\rm m},i}
&=& \frac{ m^2_{{\rm m},i}(E)} {2E},
\ea
where
$x_i$ is the position of a Lagrangian grid point of eigenstate $i$,
$v_{{\rm m},i}$ and $m^2_{{\rm m},i}$ are the group velocities and eigenvalues, respectively, in matter,
$k$ is the momentum, and
$\Lambda_{{\rm m},ij} =\Lambda_{{\rm m},i} \delta_{ij}$.
The local effective mixing matrix $U_{\rm m}$ satisfies the equation
\be
U_{\rm m}^\ast(x) \Lambda_\mathrm{m}(x) U_{\rm m}^T(x) =
U^\ast \Lambda U^T + \sqrt{2} G_\mathrm{F} n_e(x) \mathrm{diag}(1,0,0),
\ee
where $\Lambda$ is the mass matrix in vacuum. See Appendix for this adopted formulation.
The perturbation formulation based on a mean field \cite{Burgess:1996mz} is not suitable for the current case because the perturbation is large and its length scale is very short.

The evolution of neutrino wave functions $n_i(x,t)$ under a rapidly changing weak potential is exactly solved. Once mixed propagation states are generated, respective states evolve differently with different group velocities. As a result, a WP separation can be observed in the derived results. This differs from analytical evaluations. Because of different eigenvalues and velocities, nondiagonal elements of the density matrix, i.e., $\rho_{jk}(x,t) =| n(x,t) \rangle \langle n(x,t) |_{jk}$ \cite{Akhmedov:2017mcc}, defined at a fixed position oscillates.

Firstly, the evolution is solved without considering small differences in the group velocity between propagation eigenstates.
Around an inhomogeneity of $n_e$, off-diagonal components of square brackets in Eq. (\eqref{eq_evo1}) are dominant (see Ref. \cite{Fantini:2018itu} for a two-flavor case). 
Near the resonance point, conversions between $n_i$ states occur. The conversions slow down after passing through the atom, and stops when the nondiagonal terms become lower than the diagonal terms in amplitude.

Neutrino propagations outside the H-resonance region are treated.
Pre-SN profiles for $\rho$, $T$, and $Y_i$ \cite{Kusakabe:2019znq} are adopted.
Effects of only two 1s electrons of O are considered, and other electrons are assumed to distribute homogeneously.
%
Using an approximate wave function from a variational theory, the 1s $n_e$ profile averaged over angle is given by
\be
n_e^{\rm O}(r)
=\frac{2}{\pi} \left( \frac{Z'}{a_B} \right)^3
\exp \left[ -2 \left( \frac{Z'}{a_B} \right) r \right],
\ee
where
$Z' =Z -5/16$, and $r$ is the distance from the center of O atom.
The average $n_e$ along the neutrino pathway is given by
$\langle n_e \rangle = n_{e{\rm b}} +\langle n_e^{\rm O} \rangle$, where
$n_{e{\rm b}} =\langle n_e \rangle (1-f_{\rm 1s})$ is the background excluding the 1s electrons, and
$\langle n_e^{\rm O} \rangle =\langle n_e \rangle f_{\rm 1s}$ is the average contribution of 1s electrons.
The average interval of O atoms on the pathway, $r_1$, satisfies
\ba
\langle n_e^{\rm O} \rangle & =& \frac{2 \int_0^{r_1/2} n_e^{\rm O}(r) dr}{r_1}
\sim \frac{2}{\pi} \frac{(Z' /a_{\rm B})^2}{r_1},
\ea

We take $\rho =2.03 \times 10^3$ g cm$^{-3}$, where $Y_e =0.499$, $\langle n_e \rangle =6.11 \times 10^{26}$ cm$^{-3}$, $Y_{\rm O} =4.55 \times 10^{-2}$,
$f_{\rm 1s}=2 Y_{\rm O}/Y_e =0.182$, and
$r_1 =1.21 \times 10^{-8}$ cm.
%
%
The average distance between background electrons is $l_e =1.56 \times 10^{-9}$ cm, where
$l_i =2 [3 /(4 \pi n_i)]^{1/3}$.
Orbitals outer than 1s of O are unstable since the orbital radii are longer than $l_e$
.

\begin{figure}[tbp]
\begin{center}
\includegraphics[width=\linewidth]{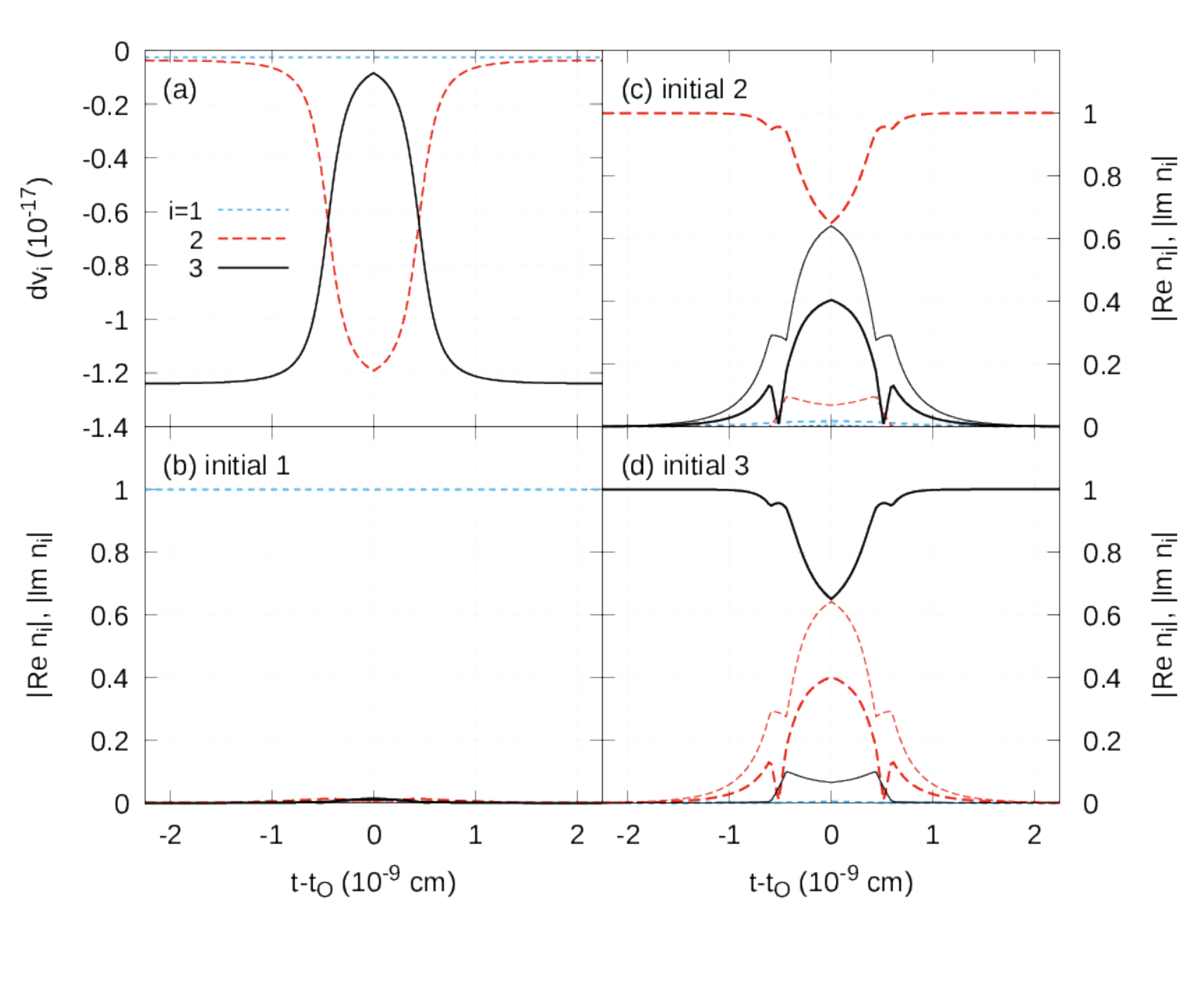}
\caption{(a) Group velocities of propagation eigenstates minus light speed, i.e., unity, as a function of propagation time from the passing through an O atom; Amplitudes of real (thick lines) and imaginary (thin lines) parts of eigenstates for cases of initial states 1 (b), 2 (c), and 3 (d). They are normalized as $\sum_j |n_j|^2 =1$. Dotted, dashed, and solid lines correspond to the eigenstates 1, 2, and 3, respectively.
  \label{fig2_eigen}}
\end{center}
\end{figure}

Figure \ref{fig2_eigen}(a) shows $dv_i =v_{{\rm m},i} -1$ as a function of time.
The group velocities change near the atom due to the large weak potential \cite{Mikheev:1987qk,Giunti:1991sx,Kersten:2013fba,Kersten:2015kio}. Fig. \ref{fig2_eigen} also shows wave functions of eigenstates $i$ for initial states 1 (b), 2, (c), and 3 (d).
For initial $n_1$ (panel b), only a tiny fraction of WP transits to $n_2$ and $n_3$, and the WP separation is minor.
For initial $n_2$ and $n_3$ cases (panels c and d), the transitions between $n_2$ and $n_3$ are significant.


It is confirmed that if group velocities were the same in this case of atomic-scale inhomogeneity, no WP separation occurs. A passing through an atom simply causes the transition of $n_i \rightarrow n_{i\neq j} \rightarrow n_i$ since the gradient of $n_e^{\rm O}(x)$ is antisymmetric around the atomic center [Eq. (\eqref{eq_evo1})].
Then, only the average $\langle n_e \rangle$ is important to determine the flavor evolution, which is consistent with earlier works \cite{Nunokawa:1996qu,Burgess:1996mz,Bamert:1997jj}.
However, due to a velocity difference, the WPs separate before a complete reproduction of the original state $n_i$ as shown below.

\section{Wave packet separations via non-adiabatic transitions}\label{sec4}

%

Second, WP evolution during passing through two O atoms is shown.
Initially, a pure $n_3$ state is assumed, and a smoothed box shape is adopted for WP in the range of $x=[x_1,x_2]$ with the length $L =x_2 -x_1 =1 \times 10^{-11}$ cm \cite{Akhmedov:2017mcc}, which corresponds to the momentum WP with the width of $\sigma_p \sim 1/(2 L) =0.99$ MeV.
Then, the height of the WP squared is $h =1/L$.
This $L$ is much shorter than the O 1s orbital radius.
The forward and rear ends are described by sigmoid functions
with the diffuseness $a =50 \Delta x$ for illustration of the effect, where $\Delta x =1\times 10^{-27}$ cm is the grid size.
The transfer equation is integrated and evolution of $x_i(t)$ and $n_i(x_i,t)$ is calculated using interpolated WP profiles.
On the neutrino path, two O atoms are placed at $r =r_1$, and $2 r_1$, respectively.
The computational range is $t =[0, 3 r_1]$.

\begin{figure}[tbp]
\begin{center}
\includegraphics[width=0.9\linewidth]{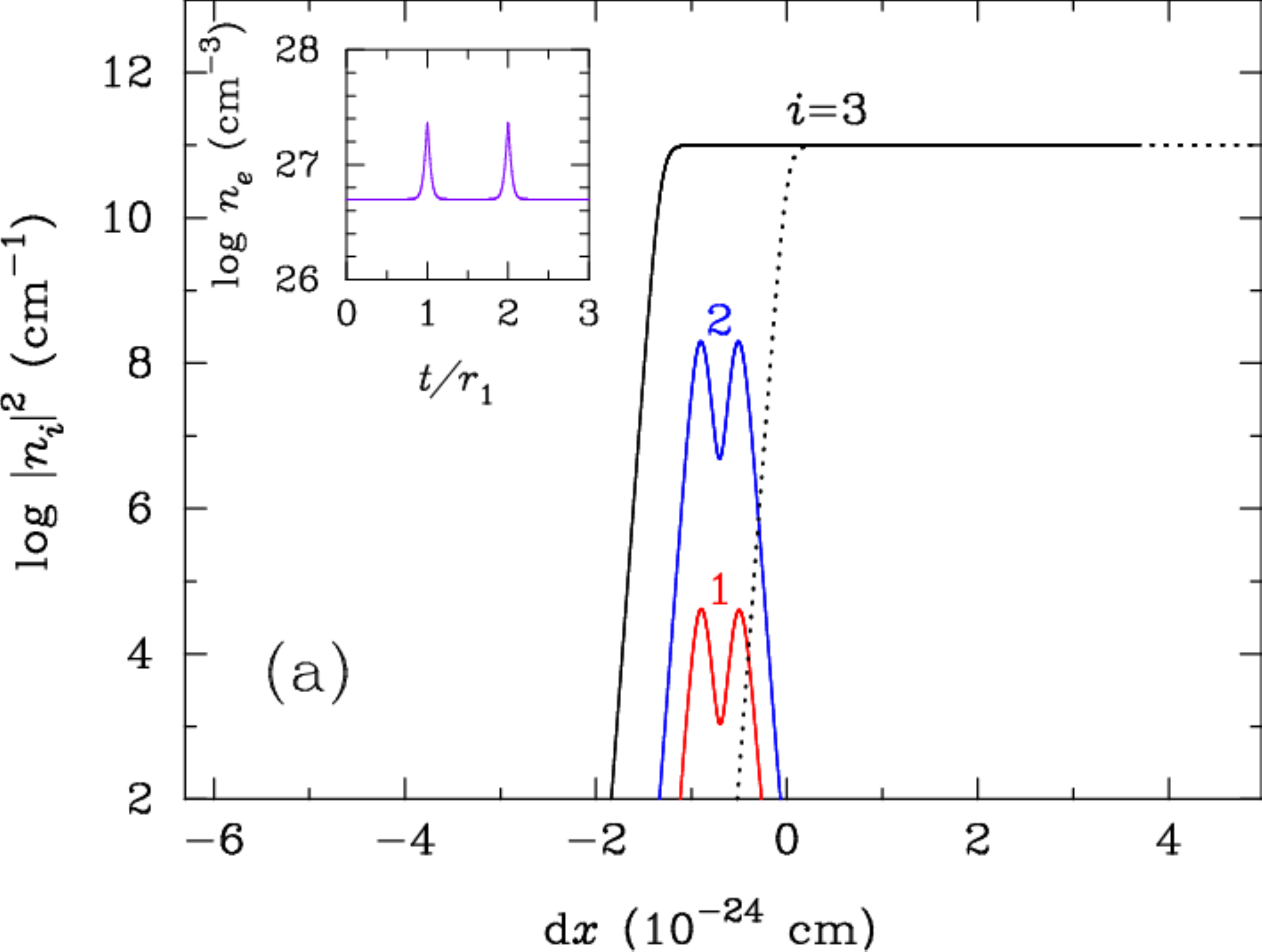}
\includegraphics[width=0.9\linewidth]{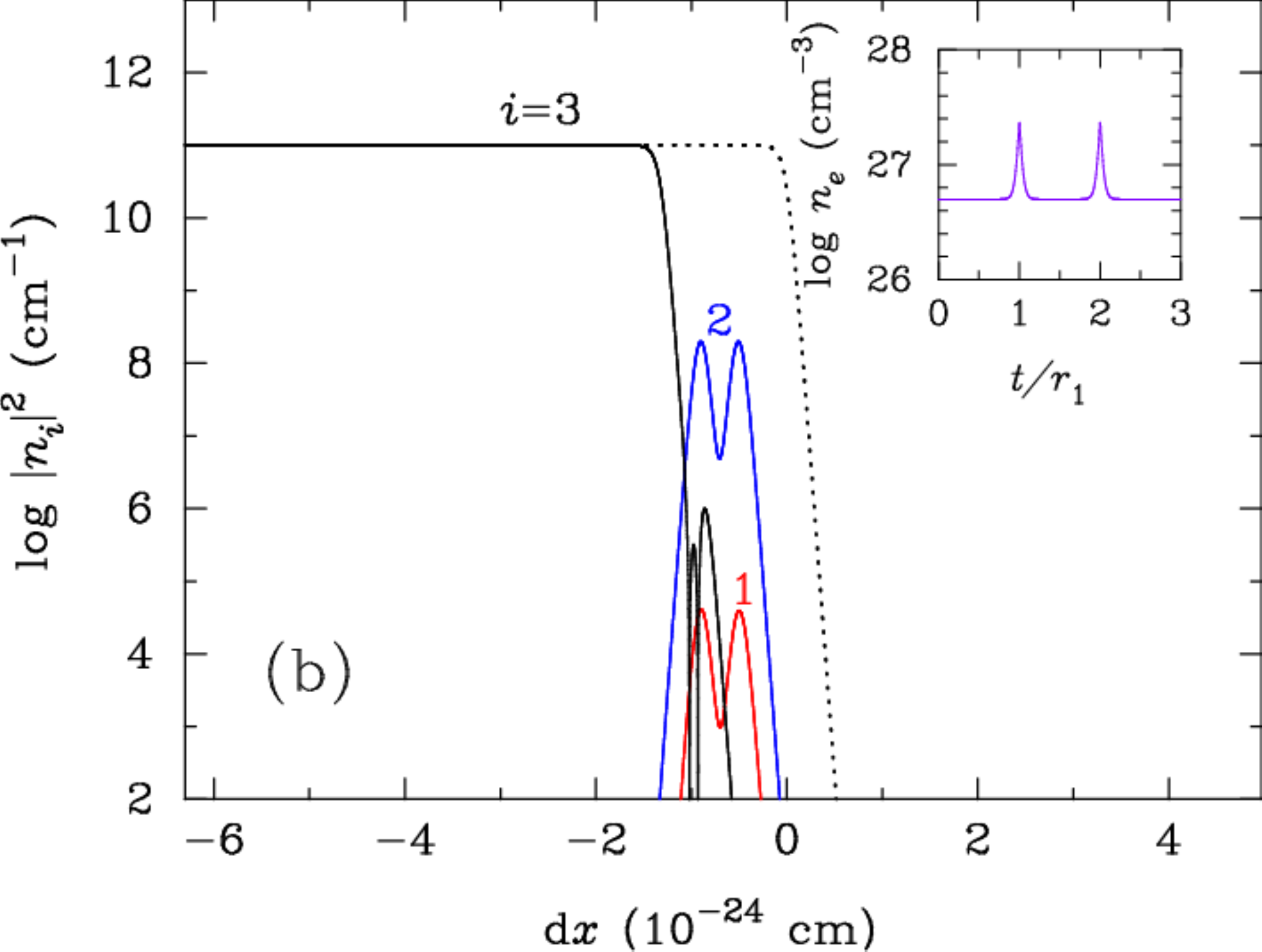}
\caption{Squares of amplitudes for neutrino propagation eigenstates versus position near the rear (a) and forward (b) ends of WPs. The initial state is purely $n_3$. The insets show the electron number density $n_e$ as a function of time. Solid and dotted lines correspond to the values at $t=3r_1$ and $t=0$, respectively. See Supplemental Material for movies of time evolution of WPs.
  \label{fig3_WP}}
\end{center}
\end{figure}

Figure \ref{fig3_WP} shows squared amplitudes of eigenstates versus position near the rear (a) and forward (b) ends of WPs.
Solid and dotted lines correspond to the final and initial values, respectively.
During the passing, conversions of states operate.
After passing through an O atom, the $n_e$ becomes homogeneous. 
Before the initial state of a pure eigenstate recovers, finite amplitudes of WPs escape from the central region.
Then, a slightly reduced amplitude of slower $n_3$ escapes from the rear end, while a finite amplitude of faster $n_2$ escapes from the forward end.
  Some significant amplitude of $n_2$ is generated also near the rear end. During the recovery of the $n_3$ state, the states $n_1$ and $n_2$ go ahead and the mixed neutrino state deviates from that of the bulk WP (Fig. \ref{fig2_eigen}). This deviation results in the final state after passing the atom that includes large mixings of $n_1$ and $n_2$.


The quantity $v_{{\rm m},i}$ only deviates from the vacuum value in a narrow region near the atom (Fig. \ref{fig2_eigen}a). The shift of WP is then approximately
$\Delta x_{ij} \approx \Delta v_{ij} t \approx \Delta m_{ji}^2 /(2E^2) r$.
For $\Delta m^2_{32}$ and $E =10$ MeV, the coherence length in vacuum is
$L_{\rm coh} =L /|\Delta v_{23}| =8.15 \times 10^5$ cm, and that in matter is of the same order of magnitude \cite{Giunti:1991sx}. 
Then, WPs separate quickly after it is emitted from a neutron star.
In the adiabatic limit, propagation eigenstates are conserved.
As a result, in an outer region of SN at $r \gg L_{\rm coh}$, states $n_i$ are well separated and there is no interference. Then, flavor transition probabilities do not oscillate \footnote{A WP of a propagation eigenstate possibly catches up with another one during propagation since the ordering of group velocities reverses (Fig. \ref{fig2_eigen}a; see also \cite{Mikheev:1987qk,Giunti:1991sx,Kersten:2013fba,Kersten:2015kio})}.
However, once the average $\langle n_e \rangle$ decreases to $n_{e,{\rm H}}$, $n_e$ is inhomogeneous due to atomic electrons. Then, propagation eigenstates effectively change and WPs separate.
Although this separation is tiny because of short passing time scale, a huge number of passings result in a significant effect.

The fraction of conversion per passing is
$f_i^{\rm WP} =h \Delta v_{23} \Delta t |N_i|^2
\sim 1.5 \times 10^{-14}~|N_i|^2$,
where
$\Delta t =r_1$ is the average time interval of passing over $^{16}$O nuclei,
$\Delta v_{23} \sim 1.23 \times 10^{-17}$, and
$|N_i|^2=|n_i|^2/h$ is the average fractional amplitude squared of $i$ at the ends of WPs.
The density scale height is $\Delta r =1.72 \times 10^9$ cm.
Then, the number of passings in this region is
$N_{\rm pass} = \Delta r /r_1
= 1.4 \times 10^{17}$.
The fraction of conversion after traveling this region is then
$N_{\rm pass} f_i^{\rm WP} =2.1 \times 10^3 |N_i|^2$.
If $N_{\rm pass} f_i^{\rm WP} \gtrsim 1$, a significant conversion of eigenstates occurs. In fact, ${\mathcal O}(10^{-3})$ of WP of initial $n_3$ separates as $n_2$ (Fig. \ref{fig3_WP}).
  Therefore, $N_{\rm pass} f_i^{\rm WP} ={\mathcal O}(1)$ and the conversion of eigenstates effectively operates for SN neutrinos.
Effects of 1s electrons of He also appear outside the L-resonance region in the He layer, and some conversions of $n_1$ and $n_2$ occur \cite{MK_future}.

  Conversions of propagation eigenstates below the average densities for the MSW resonances leads to rather equal probabilities of finding three flavors of neutrinos. This drastically changes the standard understanding of SN neutrinos. Signals of neutrinos from the next SN in or near the Galaxy are affected by this conversion mechanism. The neutrino energy spectrum and SN relic neutrino background \cite{Mirizzi:2015eza} contain signatures of the flavor conversions, and neutrino event rates in neutrino detectors can probe the signatures.
  In addition, some rare nuclei such as $^7$Li, $^{11}$B, $^{138}$La and $^{180}$Ta are produced by neutrino reactions in SN nucleosynthesis \cite{Woosley:1989bd}. Yields of light rare nuclei $^7$Li and $^{11}$B \cite{Yoshida:2006qz} as well as heavy nuclei including short-lived nuclei \cite{Ko:2020rjq} depend on the neutrino mass hierarchy. Those nuclear yields are then also a probe of the flavor conversions induced by atomic electrons in stars.

\paragraph{Sun}
The published AGSS09met model \cite{Vinyoles:2016djt} \footnote{https://www.ice.csic.es/personal/aldos/Solar\_Data.html} is adopted.
The density is $\rho \sim 100$ g cm$^{-3}$ over the range from the center to $r \sim 0.1 R_\odot$ with $R_\odot =6.96 \times 10^{10}$ cm \cite{Zyla:2020zbs}.
This is larger than the $\rho_{\rm cr}$ value of He, but smaller than that of O.
Then, most electrons distribute homogeneously, and some electrons are localized around O atoms.

%
For example, at $\rho =10$ g cm$^{-3}$, He 1s states are unstable marginally (Tab. \ref{tab1}).
There, the O mass fraction is $X_{\rm O}=6.42 \times 10^{-3}$, or $n_{\rm O} =2.43 \times 10^{21}$ cm$^{-3}$.
The average interval of O is
$r_1 =2.78 \times 10^{-4}$ cm, and
the density scale height is $\Delta r =6.23 \times 10^9$ cm.
Solar neutrinos produced in the $pp$ reaction and $^7$Be and $^8$B decays have WP widths of $L \sim 10^{-7}$ cm \cite{deHolanda:2003nj}.
Then,
$N_{\rm pass} = 2.24 \times 10^{13}$, and
%
$N_{\rm pass} f_i^{\rm WP} =2.34 (E/1~{\rm MeV})^{-2} |N_i|^2$.
Typically, this fraction is less than unity excepting low $E$ and high $|N_i|^2$.
Therefore, this loss is unimportant for solar neutrinos.

\paragraph{Earth}
The radius is $R_\oplus =6.38 \times 10^8$ cm, the mass $M_\oplus =5.97 \times 10^{27}$ g \cite{Zyla:2020zbs}, and the average density $\rho_\oplus =5.50$ g cm$^{-3}$.
The average $n_e$ is below $n_{e,{\rm H}}$ and $n_{e,{\rm L}}$ for $E \gtrsim 10$ MeV, and
slightly higher than the critical density for H.
However, heavier elements Fe, O, Si, and Mg dominate in composition (e.g., \cite{1995ChGeo.120..223M,2010E&PSL.293..259J}), and electrons in those 1s states cause inhomogeneities in $n_e$.
The O mass fraction of bulk Earth is $X_{\rm O}=0.310$ \cite{2010E&PSL.293..259J}, and
$\langle n_{\rm O} \rangle =6.41 \times 10^{22}$ cm$^{-3}$,
$r_1 =1.05 \times 10^{-5}$ cm, and 
$\langle n_e \rangle= 1.65 \times 10^{24}$ cm$^{-3}$.

Atmospheric neutrinos have a wide range of energy spectrum \cite{Richard:2015aua,Abusleme:2021paq}. They are produced via weak decays, and have wide WPs of $L \sim 20 (E/1~{\rm GeV})^{-2}$ cm \cite{Minakata:2012kg}.
They propagate in the Earth on a length scale of
$\Delta r \sim R_\oplus$, which leads to
$N_{\rm pass} = 6.09 \times 10^{13}$.
The total loss fraction of WP is
$N_{\rm pass} f_i^{\rm WP} =1.20 \times 10^{-15}~|N_i|^2$.
The effect of atomic electrons on atmospheric neutrinos is negligibly small.

\section{Summary}\label{sec4}
A mechanism for conversions of neutrino propagation eigenstates by atomic electrons is suggested.
When SN neutrinos propagate through the stellar O-layer below the MSW H-resonance density, vast numbers of non-adiabatic transitions occur by an inhomogeneous $n_e$, differently from adiabatic transitions expected with a mean $\langle n_e \rangle$ profile.
Because of a fluctuating electron potential, transitions between propagation eigenstates occur, and
WP separations effectively proceed.
SN neutrino spectra we will observe would be rather flavor-independent.
The separation is insignificant for solar and atmospheric neutrinos, and no change is required for neutrino parameters deduced until now.

\appendix

\section{Local weak potential in inhomogeneous electron background}\label{sec1}

Equations of motion from the effective Lagrangian for neutrinos propagating in nonrelativistic and unpolarized matter lead to the equation \cite{Halprin:1986pn,Akhmedov:2020vua}
\be
\left[ \vec{\nabla}^2 + E^2 -MM^\dagger - 2 E V(x) \right] \phi(x) =0,
\ee
where
$\phi$ is the two-component left-handed neutrino field,
$M$ is the neutrino mass matrix,
$E$ is the neutrino energy, and
$V(x)$ is the background potential.
In this equation, spatial component of $\vec{V}$ has been neglected due to nonrelativistic background particles, and $\vec{\nabla} V$ term has been neglected since the term of $2 E V(x)$ is larger. Also, higher order terms in $V/E$ and $MM^\dagger/E^2$ have been neglected.
Then, neutrinos propagating in the positive direction of the $x$-axis reduces  \cite{Akhmedov:2020vua} to
\be
i \frac{d}{dx} \phi(x) =
\left[ -E + \frac{MM^\dagger}{2E} + V(x) \right] \phi(x).
\ee
This is in the same form as that used in normal calculations for neutrino oscillation with averaged potential $V$ over macroscopic scale centered at $x$ \cite{Akhmedov:2020vua}.

The average charged-current (CC) potential for an electron neutrinos by electrons has been given under the assumption of a homogeneous and isotropic gas of unpolarized electrons \cite{Langacker:1982ih,Giunti:1990pp,Giunti:2007ry}. By extending the derivation to a general inhomogeneous case, a local CC potential is derived as follows. The effective CC Hamiltonian corresponding to the $W$ boson exchange between $e^-$ and $\nu_e$ is described by
\be
\mathscr{H}_\mathrm{eff}^\mathrm{(CC)}(x) =
\frac{G_\mathrm{F}}{\sqrt{2}} \left[
  \bar{\nu}_e(x) \gamma^\rho ( 1- \gamma^5 ) e(x)
  \right]
\left[
  \bar{e}(x) \gamma_\rho (1 - \gamma^5) \nu_e(x)
  \right].
\ee
where
$G_\mathrm{F}$ is the Fermi constant,
$\gamma^\rho$ is the $\gamma$ matrices,
$\gamma^5 \equiv i\gamma^0 \gamma^1 \gamma^2 \gamma^3$, and
$e(x)$ and $\nu_e(x)$ are the spinors of electron and electron neutrino, respectively.
The local effective Hamiltonian in the inhomogeneous electron background in the rest frame of the medium is given by
\ba
\mathscr{H}_\mathrm{eff}^\mathrm{(CC)}(x) &=&
\frac{G_\mathrm{F}}{\sqrt{2}}
\bar{\nu}_e(x) \gamma^\rho ( 1- \gamma^5 ) \nu_e(x)
\int d^3p_e f(E_e, x) \nonumber \\
&& \times \frac{1}{2} \sum_{h_e = \pm 1}
\langle e^-(p_e,h_e) |
\bar{e}(x) \gamma_\rho (1 - \gamma^5) e(x)
| e^-(p_e,h_e) \rangle,
\ea
where
$\bfp_e$ is the background electron momentum,
$f(E_e, x)$ is the distribution function of background electron energy $E_e$, and one-electron state is defined as
\be
| e^-(p_e,h_e) \rangle =
\frac{1}{2 E_e V} a_e^{(h_e) \dagger}(p_e) |0 \rangle,
\ee
where $a_e^{(h_e) \dagger}(p_e)$ is the creation operator of electrons with helicity $h_e$ and momentum $p_e$, and $|0\rangle$ is the ground state. For the background a finite normalization volume $V$ is introduced, and taken as small as possible in the end.
We note that compared to the homogeneous background case, the distribution function of electron has an explicit dependence on the position, and a thermal distribution for free electrons parameterized with a temperature has not been assumed here.
Four momenta and helicities of electrons before and after the scattering are identical since the interaction contributes coherently to the neutrino potential when it does not change the states of medium.

The distribution function for electron is normalized by
\be
\int d^3p_e f(E_e, x) =n_e(x) V,
\ee
where
$n_e(x)$ is the electron number density of the background in a volume $V$.

The average over helicities of electron matrix element is given \cite{Langacker:1982ih,Giunti:1990pp,Giunti:2007ry} by
\be
\frac{1}{2} \sum_{h_e = \pm 1}
\langle e^-(p_e,h_e) |
\bar{e}(x) \gamma_\rho (1 - \gamma^5) e(x)
| e^-(p_e,h_e) \rangle
=
\frac{p_{e\rho}}{E_e V}.
\ee
Then, it follows that
\be
\mathscr{H}_\mathrm{eff}^\mathrm{(CC)}(x) =
\frac{G_\mathrm{F}}{\sqrt{2}}
\frac{1}{V}
\int d^3p_e f(E_e, x)
\bar{\nu}_e(x) \frac{\gamma^\rho p_{e\rho}}{E_e} (1 - \gamma^5) \nu_e(x).
\ee
Under the assumption of isotropic electron momenta, the integral part is reduced to
\ba
\int d^3p_e f(E_e, x)
\frac{\gamma^\rho p_{e\rho}}{E_e}
&=&
\int d^3p_e f(E_e, x)
\left (\gamma^0 -\frac{\vec{\gamma} \cdot \vec{p}_{e}}{E_e} \right) \nonumber \\
&=& n_e(x) V \gamma^0,
\ea
where the second term on the right-hand side becomes zero after integration due to isotropic momentum distribution. Then, we obtain the local CC Hamiltonian given by
\ba
\mathscr{H}_\mathrm{eff}^\mathrm{(CC)}(x) &=&
V_\mathrm{CC}(x)
\bar{\nu}_{e\mathrm{L}}(x) \gamma^0 \nu_{e\mathrm{L}}(x) \\
V_\mathrm{CC}(x) &=&
\sqrt{2} G_\mathrm{F} n_e(x),
\ea
where
$\nu_{e\mathrm{L}}(x)=[(1-\gamma^5)/2] \nu_{e}(x)$ is the left-handed $\nu_e$.
Therefore, electron neutrinos propagating in inhomogeneous electron background are affected by local electron potential on their pathways.

Neutrinos traveling in matter are refracted \cite{Wolfenstein:1977ue} via weak interactions of particles in the matter. If the propagation eigenstates and the flavor eigenstates were identical, the energy is related to the momentum $k_\mathrm{m}$ in matter \cite{Fukugita:2003en} simply by
\be
E = \sqrt{k_\mathrm{m}(x)^2 +m^2} +V(x).
\ee
The refractive index $n^\mathrm{r}$ is defined by
\be
\Psi(x) =\exp\left( i n^\mathrm{r}(x) \bfk \cdot \bfx - i E t \right),
\ee
where $\Psi$ is the wave function, 
$\bfk$ is the wave vector in the vacuum related by $k^2 +m^2 =E^2$. From $k_\mathrm{m}(x) =n^\mathrm{r}(x)k$, the refractive index satisfies
\be
n^\mathrm{r}(x) = 1 - \frac{EV(x)}{k^2}.
\ee
In reality, the propagation eigenstates of neutrinos are different from the flavor eigenstates. As a result, the refractive index for propagation eigenstates $n^\mathrm{r}_i$ is given \cite{Mikheev:1987qk} by
\ba
n^\mathrm{r}_i(x) &=& \frac{v^\mathrm{p}_i}{v^\mathrm{p}_{{\rm m},i}(x)} \\
v^\mathrm{p}_{{\rm m},i}(x) &=& \frac{E_i(x)}{k} =1 +\frac{\Lambda_{{\rm m},i}(E; x)}{k} \\
\Lambda_{{\rm m},i}(E; x)
&=& \frac{ m^2_{{\rm m},i}(E; x)} {2E},
\ea
where
$v^\mathrm{p}_i$ and $v^\mathrm{p}_{{\rm m},i}$ are the phase velocities of propagation eigenstates $i$ in vacuum and matter, respectively,
$E_i$ and $m^2_{{\rm m},i}$ are the eigenenergy and the mass squared, respectively in matter.



\bibliography{ref1}



\end{document}